\documentclass[12pt]{article}
\usepackage{epsfig}
\usepackage{graphicx}

\newcommand{\be}{\begin{equation}}
\newcommand{\ee}{\end{equation}}
\def\({\left (}
\def\){\right )}
\def\[{\left [}
\def\]{\right ]}
 \def\al{\alpha}
 \def\ga{\gamma}
 \def\ra{\rightarrow}
 \def\vp{\varphi}
 
\begin{document}

\begin{titlepage}
\bigskip
\rightline{}
\rightline{hep-th/0506166}
\bigskip\bigskip\bigskip\bigskip
\centerline {\Large \bf {Tachyon Condensation and Black Strings}}
\bigskip\bigskip
\bigskip\bigskip

\centerline{\large Gary T. Horowitz}
\bigskip\bigskip
\centerline{\em Department of Physics, UCSB, Santa Barbara, CA 93106}
\centerline{\em gary@physics.ucsb.edu}
\bigskip\bigskip

\begin{abstract}
We show that under certain conditions, closed string tachyon condensation produces a topology changing transition from black strings to Kaluza-Klein ``bubbles of nothing."  This  can occur when the curvature at the horizon is much smaller than the string scale,  so the black string is far from the correspondence point when it would make a transition to an excited fundamental string.  This provides a dramatic new endpoint to Hawking evaporation.  A similar transition occurs for  black $p$-branes, and can be viewed as a nonextremal version of a geometric transition. Applications to  AdS black holes and the AdS soliton are also discussed.
\end{abstract}

\end{titlepage}

\baselineskip=16pt
\setcounter{equation}{0}
\section{Introduction}
It is well known that strings wound around a circle with antiperiodic  fermions become tachyonic when the size of the circle shrinks below the string scale \cite{Rohm:1983aq}. In a recent paper \cite{Adams:2005rb} strong evidence was given that when these tachyons are localized, the outcome of this instability is that the circle pinches off. This produces a topology changing transition in which handles can disappear or space can become disconnected.

 In this paper we explore applications of this tachyon instability to black strings and branes. A charged black string wrapped around a circle causes the size of the circle to shrink to zero at the singularity. Hence it must reach the string scale at some radius which can lie inside or outside the horizon. If the circle has antiperiodic boundary conditions for fermions, then the tachyon instability will set in at this point. If the radius is inside the horizon, this occurs along a spacelike surface and the evolution past this point is no longer governed by general relativity. The spacelike singularity is replaced by the physics of the tachyon condensation. This has recently been investigated in the context of cosmology in \cite{MS} and will be applied to black strings elsewhere \cite{HMS}. 
 
  Here we focus on the other possibility that
  the circle reaches the string scale outside the horizon.  This case is more analogous to the topology changing transitions discussed in \cite{Adams:2005rb}, and tachyon condensation will cause the circle to pinch off. The resulting spacetime no longer has a horizon and describes a Kaluza-Klein (KK) bubble. KK bubbles have previously been discussed in terms of an instability of the KK vacuum \cite{Witten:1981gj}. A nonperturbative quantum gravity process can cause flat space cross a circle to nucleate a ``bubble of nothing" which rapidly expands.  For circles  larger than the Planck scale, this is a very rare process.  Here we have a new mechanism for forming such bubbles starting with a charged black string. In fact, one can argue that this is an inevitable consequence of Hawking radiation. Even if one starts with a charged black string with the wrapped circle larger than the string scale everywhere outside the horizon, Hawking radiation will cause the black string to approach extremality.\footnote{Winding modes are light  near the horizon, but massive at infinity. So they are effectively confined and one cannot radiate away much charge.}  The size of the circle at the horizon will shrink and eventually reach the string scale.  

 It was argued in \cite{Horowitz:1996nw} (see also \cite{Susskind:1993ws})  that charged black strings would Hawking radiate down to a correspondence point where the curvature at the horizon reaches the string scale. At this point, the black string makes a transition to an excited state of a fundamental winding string.  We will see that (with antiperiodic boundary conditions around the circle) the tachyon instability can set in before the correspondence point is reached.  As soon as the size of the circle reaches the string scale outside the horizon, the circle pinches off, the horizon is lost, and one forms a KK bubble.
 
 Unlike most previous discussions of KK bubbles, these bubbles have charge, since the charge on the black string is conserved during tachyon condensation. (Charged bubbles in anti de Sitter space (AdS) were discussed in \cite{Biswas:2004xc}.) We will present  exact solutions of static charged bubbles and also discuss more general bubble solutions. Both the initial black string and final bubbles can be well described by supergravity. Only the transition itself requires stringy physics. It turns out that some black strings cannot decay to a static bubble. The resulting bubble will expand outward. This provides a  dramatic (and potentially catastrophic) endpoint to Hawking evaporation.

A similar phenomenon occurs for black $p$-branes with RR charge. If one wraps one of these branes  around an appropriate circle, tachyon condensation will again produce a charged bubble. This is a nonextremal analog of what has been observed for D-branes.  It has been shown that under certain conditions, a large number of branes wrapped around a cycle have a dual description in which the branes are gone and replaced by fluxes \cite{Gopakumar:1998ki,Vafa:2000wi}. This has become known as a geometric transition. Since the bubble geometry contains flux but no branes,   tachyon condensation provides a  nonextremal version of this geometric transition.

A circle with antiperiodic fermions has a one-loop Casimir energy which will cause it to contract. So it is driven toward the tachyon instability and is not really suitable for boundary conditions at infinity. However, it was shown in \cite{Saltman:2004jh}  that one can stabilize such circles by adding suitable fluxes and branes. In particular, perturbatively stable compactifications on Riemann surfaces were found. Around some of the one-cycles, fermions must be antiperiodic.  We will discuss asymptotically flat black strings with the understanding that they are good approximations to black strings wrapping circles of this type. 

In the next section we consider black strings with fundamental string winding charge and discuss the transition to  bubbles of nothing. In section 3 we consider the tachyon instability for two charged black strings, BTZ black holes (which are just their near horizon limit), and  general bubble initial data. Section 4  contains applications to $p$-branes and other AdS black holes. The final section contains some concluding remarks and open questions.

\setcounter{equation}{0}
\section{ Black strings with F1 charge}

  The simplest black string is the product of a Schwarzschild black hole and a line. If one boosts this solution and applies T-duality, one obtains a black string with F1 (fundamental string) winding charge. In $D=n+4$ dimensions, the (string frame) metric   is
 \be\label{bsmetric}
  ds^2 = H_1^{-1}(r) [-f(r) dt^2 + dx^2] + f^{-1}(r) dr^2 + r^2 d\Omega_{n+1} 
  \ee
  where
 \be\label{Hfdef}
 H_1(r) = 1+ {r_0^n\sinh^2\al\over r^n}, \qquad f(r) = 1-{r_0^n\over r^n}
 \ee
 The two-form potential and dilaton are
 \be
 B_{xt} = {r_0^n \sinh 2\al\over 2(r^n+r_0^n\sinh^2\al)} , \qquad e^{-2\Phi} = H_1
 \ee
 The solution depends on two free parameters,  $\al$ and the horizon radius $r_0$, which determine the mass and charge:
\be\label{bsmass}
M =C_n r_0^n\[ {n+2\over n} +  \cosh2\al\]  
\ee
where $C_n$ is a dimension dependent constant, and the winding number is proportional to
\be\label{bscharge}
Q= r_0^n \sinh 2\al
\ee
Note that the dilaton goes to weak coupling near the singularity. The above solution assumes $\Phi\ra 0$ asymptotically, but it can be shifted to have arbitrarily weak string coupling everywhere.

 Let us periodically identify $x$ with period $L_1$ and impose antiperiodic boundary conditions for fermions. Then the length of this circle monotonically shrinks from $L_1$ at infinity to zero at the singularity. So it will reach the string scale when $H_1(r) = L_1^2/l_s^2$, which can either lie inside or outside the horizon.   
 We focus on the  case that
  the circle reaches the string scale near the horizon.  As mentioned in the introduction, this is a natural consequence of Hawking evaporation. The analysis in  \cite{Adams:2005rb} then shows that  tachyon condensation will cause the circle to pinch off. The resulting spacetime no longer has a horizon and describes a Kaluza-Klein bubble\footnote{There is no tachyon instability in the KK bubble even though the size of the circle shrinks to zero, since a tachyon only  exists  when the circle size does not change rapidly with radial distance.}.   The circle will be string scale at the horizon if  $\cosh\al = L_1/l_s$.  To avoid tachyon condensation at large radius,  we want $L_1 \gg l_s$, which implies large $\al$.  However
  the curvature at the horizon is of order $1/r_0^2$ which will still be much smaller than the string scale if $r_0 \gg l_s$.   

One question that immediately arises is what happens to the black string entropy? The initial black string can have a large entropy (computed from the horizon area in the Einstein metric)
\be
S= {r_0^{n+1} L_1 \Omega_{n+1}\over 4 G_D}\cosh\al
\ee 
where $\Omega_{n+1}$ is the area of a unit $S^{n+1}$.
  Since a KK bubble has no intrinsic entropy, tachyon condensation must produce some radiation (or excited strings) in addition to the bubble to account for the black string entropy. Thus the mass of the final bubble must be less than the mass of the original black string.  We will see in the next section that this is indeed the case. Of course, this does not answer the question of how the information gets out. That is undoubtedly related to how the tachyon condensation inside the horizon matches on the bubble formation outside the horizon when the black string evaporates \cite{HMS}.

Let us try to describe the resulting bubbles in more detail.
 These bubbles must have nonzero F1 charge,  since charge is conserved in the process of tachyon condensation.
It turns out to be easy to write down exact supergravity solutions describing static charged bubbles. They can be obtained from the charged black string above by a simple analytic continuation  $t\rightarrow i\chi$, $x\rightarrow i t$. The result is
\be\label{F1bubble}
 ds^2 = H_1^{-1}(r) [- dt^2 + f(r) d\chi^2] + f^{-1}(r) dr^2 + r^2 d\Omega_{n+1} 
 \ee
The dilaton  and  $B_{\mu\nu}$ are both unchanged. To avoid conical singularities at $r=r_0$ we must periodically identify $\chi$ with period
\be\label{period}
L_b =  {4\pi\over f'(r_0)} H^{1/2}(r_0) = {4\pi\over n} r_0 \cosh\al
\ee
Since $B_{\mu\nu}$ is unchanged, so is $Q$.  However, there is no longer a singularity to act as a source of the charge. The charge is now  a result of flux on the noncontractible $S^{n+1}$.  The radius of the bubble, $r_0$,  agrees with the radius of the $S^{n+1}$ on the horizon. This is  expected since the tachyon condensation which pinches off the circle should not affect the area of the transverse sphere. So it is tempting to conjecture that tachyon condensation of the black string produces this static bubble. However there is a problem. The length of the circle at infinity must also be unchanged. If we start with $L_1 = l_s\cosh\al$  and set $L_1=L_b$, then (\ref{period}) implies that $r_0$ is of order the string scale. The only charged black strings which can decay to a static bubble are those where the curvature at the horizon is of order the string scale.  For the black strings we are interested in, the resulting bubble cannot be static. We will discuss these dynamical bubbles in the next section, but for later reference, let us discuss some properties of the static charged bubbles.

Since $r_0$ is a free parameter in (\ref{F1bubble}), static charged bubbles come in all sizes. However its clear from (\ref{period}) that the size of the bubble is always less than $L_b$.
One can understand the existence of these static  bubbles intuitively as follows. As we will review below, in the absence of matter  bubbles exist  in all sizes. Bubbles which are smaller than the size of the circle at infinity contract, and larger ones expand. Putting flux on the sphere causes it to expand, but in some cases one can start with a vacuum bubble that would normally contract, and add the flux to obtain a static solution. However
one has a bound on how much flux can be carried by a static bubble
\be\label{bound}
{Q\over L_b^n} \propto  {\sinh\al\over \cosh^{n-1}\al} 
\ee
This is bounded from above for all $n>1$, i.e., in more than five dimensions.

We now comment on the perturbative stability of the static charged bubbles. While not rigorous, the following argument gives a good idea when the static bubble is likely to be classically unstable. This  problem is directly related to the stability of the black strings as follows \cite{Sarbach:2004rm}.  We can decompose perturbations of the static bubble in terms of $e^{ik\chi}$. If the bubble has an unstable mode for some $k$, then it probably has one for $k=0$. But under analytic continuation, an $\chi$-independent mode which grows like $e^{\Omega t}$ translates into a static perturbation of the black string with spatial dependence $e^{i\Omega x}$.  There is a long history of studying perturbations of black strings starting with Gregory and Laflamme \cite{Gregory:1993vy}. Short wavelength modes are always stable, but long wavelength modes are sometimes unstable. If they are, there must be a critical wavelength in which the perturbation is static. Conversely, a static perturbation is a good indication that longer wavelength perturbations will be unstable. The Gubser-Mitra conjecture \cite{Gubser:2000ec} states that (infinitely long) charged black strings will be unstable if and only if their specific heat is negative.

 Putting these results together, we see that the stability of static bubbles (with large $L_b$) depends on the sign of the specific heat of the corresponding black string.
This turns out to be dimension dependent. In $D=5,6$, the specific heat is always negative, while in $D> 6$, the specific heat is positive in the near extremal limit. The net result is that for  $D=5,6$ and large $L_1$, there is always a static perturbation of the black string, and hence an unstable perturbation of the corresponding  bubble. In higher dimensions,  the  static bubbles are perturbatively stable or unstable depending on how close to extremality the corresponding black string is.  We will argue below that in all cases, these bubbles are  nonperturbatively unstable.

Notice that the extremal limit ($r_0 \ra 0, Q$ fixed) of the black string (\ref{bsmetric}) and static bubble (\ref{F1bubble}) are exactly the same. They both reduce to the field of a wound string. (Of course the curvature gets large in both cases in this limit.) So we can think of them, formally,  as different resolutions of the winding string singularity. 

\setcounter{equation}{0}
\section{Two Charged Black Strings}

 If we compactify on $T^4$ and consider the above black string  in $D=6$, then we can easily add a NS five-brane charge. The (string metric) for the resulting  black string is
 \be\label{F1NS5bs}
  ds^2 = H_1^{-1}(r) [-f(r) dt^2 + dx^2] + H_5(r)[ f^{-1}(r) dr^2 + r^2 d\Omega_{3}] 
  \ee
where $H_1(r)$ and $f(r)$ are given  in (\ref{Hfdef})  (with $n=2$) and
 \be
 H_5(r) = 1+ {r_0^2\sinh^2\ga\over r^2}
  \ee
  The dilaton is now
  \be
e^{-2\Phi} = {H_1\over H_5}
\ee
and the mass is
\be
M= C r_0^2 (1+ \cosh2\al+\cosh2\ga)
\ee
with $C=\pi L/8G_6$.
The addition of the five-brane increases the area of the horizon, but does not change the fact that the $x$ circle shrinks to zero size at the singularity.  So our previous discussion still applies and if the circle has length $L=l_s\cosh\al$ asymptotically, the size of the circle at the horizon is of order the string scale, and  we expect a transition to a  bubble. The horizon curvature will be less than the string scale provided $r_0 \cosh\al \gg l_s$. Note that there is no need to require $r_0 >l_s$.

 There are three advantages of adding the five-brane charge which we discuss in this section. First, if $Q/L^2$ is small enough, these black strings can decay to static bubbles. Second, if we set $\al=\ga$ the dilaton decouples. This will make it easier to discuss dynamical bubbles. Third, we can take a near horizon limit and obtain results about BTZ black holes in $AdS_3$. 

\subsection{Static bubbles}

Let us set $\al=\ga$, so $ H_1(r) =H_5(r) \equiv H(r)$.  The integer normalized charges are proportional to 
\be
Q=r_0^2\sinh 2\al
\ee
Tachyon condensation will produce a  bubble with both F1 and NS5 charge.\footnote{Actually, since the dilaton vanishes, the above string metric is equivalent to the Einstein metric, so S-duality implies the same metric describes a black string with D1-D5 charges.}  A family of static  bubbles with these charges can again be obtained by double analytic continuation:
\be\label{F1NS5bubble}
 ds^2 = H^{-1}(r) [- dt^2 + f(r) d\chi^2] + H(r)[ f^{-1}(r) dr^2 + r^2 d\Omega_{3} ]
 \ee
 Regularity at $r=r_0$ requires that $\chi$ be periodically identified with period
 \be
 L=  2\pi r_0 \cosh^2 \al
 \ee
 If we insist that $r_0$ and $\al$ are both unchanged during tachyon condensation, we again run into a problem. Setting $L= l_s\cosh\al$, we find $r_0 \cosh \al =l_s$ so the radius of the bubble is again the string scale. This would imply that black strings with small curvature at the horizon cannot decay to static bubbles. But there is no justification for
 insisting that $r_0$ and $\al$ are both unchanged. Physically, the only requirements are that the size of the horizon equal the size of the bubble, and that $Q$ and $L$ are conserved. If we let $\tilde r_0$ and $\tilde \al$ denote the corresponding quantities for the bubble, all three conditions are satisfied (for large $\al$, $\tilde \al$) if
 \be
 r_0 e^\al = \tilde r_0 e^{\tilde \al}, \qquad l_s e^\al = \tilde r_0 e^{2\tilde \al}
 \ee
 These two equations imply $r_0 e^{\tilde \al} =l_s$. If we want the size of the horizon much larger than  the string scale, $r_0e^\al \gg l_s$, we  can simply choose $\al \gg \tilde \al$.
 This freedom is not available in the single charged case, since the requirement that the size of the horizon agree with the size of the bubble requires that $r_0=\tilde r_0$.
 
 However there is still a constraint on which black strings can decay to static bubbles.
The ratio of the charge $Q$ to the size of the circle at infinity is now
\be\label{twocharged}
{Q\over L^2} \propto {\sinh\al\over \cosh^3\al}
\ee
which is again bounded from above. So  only some of the charged black strings can become
static charged bubbles after tachyon condensation. To see what happens to the others, we need to consider more general bubble solutions.

\subsection{General bubble initial data}

We will consider time symmetric initial data. As noted above, the advantage of adding the five-brane is that we can set the dilaton to zero, and only consider the metric and three-form.  To get started, we  will set the charge to zero and just consider vacuum initial data. Consider a spatial metric of the form
\be\label{bubbleid}
 ds^2 = U(\rho) d\chi^2 + { d\rho^2\over U(\rho) h(\rho)} + \rho^2 d\Omega_3
 \ee
Since we set the time derivative of the metric to zero, the only constraint for the vacuum initial data is the vanishing of the scalar curvature:
\be\label{scalar}
{}^5R= -{1\over 2\rho^2} [ -12 + h'(\rho^2 U' + 6\rho U) + 2h (\rho^2 U'' + 6\rho U' + 6U)]=0
\ee
Since there are two functions and only one constraint, there are many solutions. We can pick $U$ arbitrarily and then solve (\ref{scalar}) for $h$. To obtain bubble initial data, we want $U$ to have a simple zero at a positive value of $\rho$. So let us choose
\be\label{defU}
U(\rho) = 1-{\rho_0^2\over \rho^2}
\ee
The resulting equation is easily solved for $h$ with the result
\be\label{defh}
h(\rho) = 1+ {b\over 3\rho^2 - 2\rho_0^2}
\ee
We thus obtain a two parameter family of  time symmetric,  asymptotically flat initial data. The metric would be  singular at $\rho^2 = 2\rho_0^2/3$, but the space smoothly pinches off at $\rho=\rho_0$ provided  that we identify $\chi$ with period
\be\label{length}
L = {2\pi \rho_0\over (1 + {b\over \rho_0^2})^{1/2}}
\ee
Since any $L$ can be obtained by adjusting the free parameter $b$, we obtain initial data describing vacuum  bubbles of arbitrary size $\rho_0$ in a space with any size circle  at infinity.

The mass of  these bubbles are
\be
M= C\[ 2\rho^2_0 - {4\pi^2 \rho_0^4\over L^2}\] 
\ee
where we have replaced the parameter $b$ by $L$.
Clearly, the mass grows with radius for small bubbles, but  reaches a maximum when $\rho_0 = L/2\pi$ and then decreases with radius and eventually becomes negative.
This shows that the positive energy theorem does not hold for our initial data. The usual spinorial proof \cite{Witten:1981mf} does not apply since that requires asymptotically constant spinors and in the bubble spacetime, all fermions must be antiperiodic around the circle at infinity.

The bubble size which maximizes the mass corresponds to initial data for a static bubble.  The static bubble is simply a product of time and the Euclidean 5D Schwarzschild metric. (This is also the double analytic continuation of a neutral black string.)  Its initial data corresponds to (\ref{bubbleid}, \ref{defU}, \ref{defh})  with $b=0$. From (\ref{length}) this implies $\rho_0 = L/2\pi$, which  indeed corresponds to the maximum of $M$. 

The initial data we have constructed is analogous to the bubble initial data in one lower dimension discussed in \cite{Brill:1991qe}. One can analytically determine whether the bubble is initially expanding or contracting \cite{Corley:1994mc}, and the full 
 time evolution  has been found numerically
 \cite{Sarbach:2003dz}. The result is that
small bubbles collapse\footnote{One might have thought that the small bubbles would collapse to (neutral, uniform) black strings, but in less than ten dimensions, the mass of the bubble is so low that the corresponding black strings would be unstable \cite{Elvang:2004iz}. Since the evolution produces trapped surfaces, it must evolve to the end state of the Gregory-Laflamme instability - which is still unknown.}
  and large bubbles continue to expand indefinitely. One example where the evolution is known analytically, is the zero mass bubble first obtained by Witten \cite{Witten:1981gj}.  We expect that the evolution of the above initial data will be similar, with small bubbles collapsing and large bubbles expanding.  The analog of Witten's bubble can be obtained by analytically continuing the 6D Schwarzschild solution. This would correspond to initial data with $U= 1- \rho_0^3/\rho^3$, $h=1$ and is not included in the above set.

Now let us add charge to these bubbles. This corresponds to putting three-form flux on the $S^3$. Intuitively, one expects that a nonzero charge will raise the energy of small bubbles, but it should have little effect on large bubbles. In particular, charged bubbles with negative energy should still exist. To construct initial data for charged bubbles, we set the three-form to $Q (\epsilon_3 + {}^*\epsilon_3)$, where $\epsilon_3$ is the volume form on a unit $S^3$. Then the constraint becomes ${}^5R = Q^2/\rho^6$. If we keep $U$ unchanged, the solution for $h$ becomes
 \be
h(\rho) = 1+ {b\over 3\rho^2 - 2\rho_0^2} - {Q^2\over 4\rho_0^2 \rho^2}
\ee 
The length of the circle at infinity becomes
\be
L = {2\pi \rho_0\over (1 + {b\over \rho_0^2} - {Q^2\over 4\rho_0^4})^{1/2}}
\ee
and the mass becomes
\be\label{bubblemass}
M = C\[  {Q^2\over 2\rho_0^2}  + 2\rho^2_0 - {4\pi^2 \rho_0^4\over L^2}\] 
\ee
The mass now diverges as the bubble size shrinks to zero. This is the familiar effect of flux stabilizing small cycles. For $L^2\gg Q$, there is a local minimum at $\rho_0^2 \approx Q/2$ and a local maximum at $\rho_0 \approx L/2\pi$ (see Fig. 1). One can easily verify that these extrema in the mass for fixed $L$ again correspond to $b=0$. They correspond to  two of our static bubbles obtained by analytically continuing the charged black string. To show this, one takes a $t=$ constant surface in the static charged bubble (\ref{F1NS5bubble}) and introduces a new radial coordinate $\rho^2 = r^2 + r_0^2 \sinh^2\al $. Then the spatial metric takes the form (\ref{bubbleid}) with $\rho_0 =r_0 \cosh \al$ and $Q^2/4\rho_0^2 = r_0^2 \sinh^2\al$. Within this family of initial data, the smaller bubble is locally stable, but the larger bubble is unstable. From the argument in the previous section, we expect the smaller bubble to be stable to all small perturbations since the $D=6$ two charged black string has positive specific heat near extremality. However, since the mass (\ref{bubblemass}) is unbounded from below, even the locally stable bubble will be nonperturbatively unstable.

%%%%%%%%%%%%%%%%%%%%%%%%%%%%%%%%
\begin{figure}[htb]
\begin{picture}(0,0)
\end{picture}
\mbox{\epsfxsize=15cm \epsffile{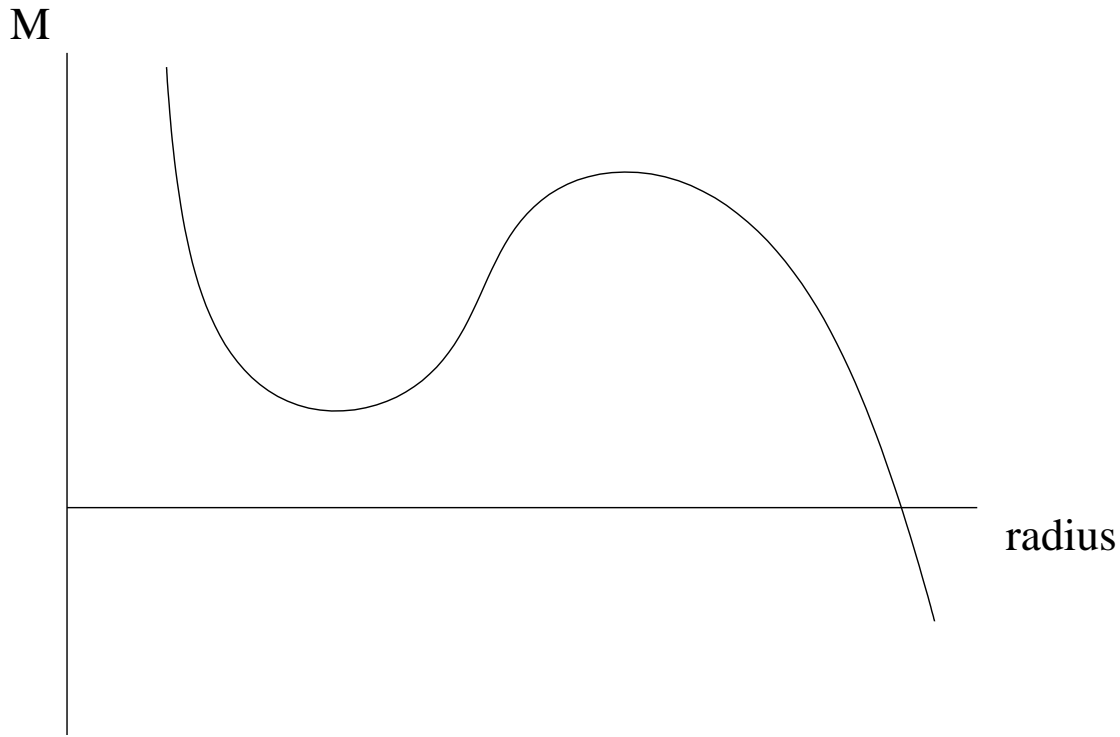}}
\caption{The mass of charged bubbles as a function of their radius, assuming $Q<L^2$.
The local minimum is at $\rho^2_0\approx Q/2$ and the local maximum is at $\rho_0 \approx L/2\pi$.}
\label{1}
\end{figure}
%%%%%%%%%%%%%%%%%%%%%%%%%%%%%%%%%%%

The key question now is which of the above initial data describes the end state of tachyon condensation of our  black strings. Since our black strings have large $\al$, their horizon radius is set by the charge. This means that the resulting  bubble must have $\rho^2_0\approx Q/2$. Hence if the initial black string has $Q \ll  L^2$, the decay will produce the static bubble. On the other hand, if 
  $L^2 <Q$,  $M$ is a monotonically decreasing function of the bubble radius $\rho_0$. There are no extrema, and no static bubbles. Tachyon condensation produces a bubble of nothing  which  expands out.\footnote{In the nonstatic case, there is no guarantee that  tachyon condensation will yield precisely one of the bubbles in our three parameter family. The above solutions are  examples of the general behavior.} 

We stated earlier that in order for entropy to increase during the process of tachyon condensation,  the mass of the bubble produced must be less than the mass of the black string. It is easy to see that this is indeed the case.  Our original black strings have a mass slightly larger than the extremal limit:   $M=C(2Q +r_0^2)$. The mass of our  bubble initial data with $\rho_0^2 = Q/2$ is always {\it less} than the extremal mass. For any $L$, $M=C(2Q - \pi^2 Q^2/L^2)$. So there is always energy left over to carry the entropy. 

One cannot stabilize an expanding bubble by wrapping a brane around it. Since the bubble has co-dimension three, the gravitational effect of a wrapped brane is similar to a point mass in three dimensions, and just produces a deficit angle. Ignoring the brane charge,  a good model of the resulting spacetime is the usual bubble with the coordinate $\chi$ periodically identified with a period smaller than usual. Clearly, the dynamics is unaffected, and the bubble continues to expand.\footnote{I thank E. Silverstein for pointing this out.} 

\subsection{$AdS_3 \times S^3$}

If we take the near horizon limit of the two charged black string (\ref{F1NS5bs}), one obtains the product of a nonrotating BTZ black hole and $S^3$. This corresponds to taking $r \ll r_0\sinh\al$, and  $r\ll r_0\sinh\ga $ so one can drop the one in the harmonic functions $H_1$ and $H_5$. Rescaling the coordinates puts the metric
into the standard BTZ form
\be
ds^2 = -\({\hat r^2 - \hat r_0^2 \over \ell^2}\) dt^2 + {\ell^2 d\hat r^2\over \hat r^2-\hat r_0^2} + \hat r^2 d\vp^2
\ee
where $\ell = r_0\sinh\ga$.
Tachyon condensation  will occur near the horizon if $\hat r_0 \sim l_s$. Since the curvature is constant and set by $\ell$, $\al'$ corrections are still small.  However, there is an important qualitative difference between this three dimensional case and the higher dimensional cases discussed above: pinching off the $\vp$ circle just produces a space with the same topology as $AdS_3$. In fact, the analog of the static bubble, obtained by double analytic continuation $t\ra i\vp$ and $\vp\ra it$ is just global $AdS_3$. This follows from the fact that it must have constant curvature and the same global topology as $AdS_3$. (It can also be checked explicitly, and holds for all values of the BTZ radius $\hat r_0$.) So tachyon condensation in this case does not lead to an instability. The BTZ black hole simply decays to $AdS_3$ with radiation. 

Note that there is no constraint analogous to the bound on $Q/L^2$ for black strings to form static bubbles in the asymptotically flat case. Here, $L$ is essentially infinite and all BTZ black holes decay to global $AdS_3$

It is a good thing that there is no instability with $AdS_3 \times S^3$ boundary conditions.
 Global $AdS_3$ corresponds to  the NS vacuum for the dual $1+1$ CFT on the boundary, so the fermions must be antiperiodic.  Hence all black holes which form from nonsingular collapse of matter in $AdS_3$ must have antiperiodic boundary conditions. If there was an analog of bubbles of nothing cutting off all space, or solutions with arbitrarily negative energy, this sector of the theory would be completely unstable. This would either indicate an unexpected instability of the dual CFT, or a problem with the AdS/CFT correspondence.   The RR ground state of the CFT corresponds to the $M=0$ black hole with periodic fermions. A black hole in this sector could Hawking radiate down to a $M=0$ black hole. However this cannot happen for a black hole formed from gravitational collapse. The black hole evaporates until $r_0\sim l_s$ and then tachyon condensation causes a transition to $AdS_3\times S^3$ with radiation. In effect, it smooths out the endpoint of Hawking evaporation.
 
 If one starts with an eternal BTZ black hole, one can impose either periodic or antiperiodic boundary conditions for the fermions. Since there are two asymptotically $AdS_3$ regions, in the latter case tachyon condensation will produce two copies of $AdS_3$. In terms of the dual CFT's, if the initial state before tachyon condensation is not a product of a state in one asymptotic CFT with a state in the other, then the final state will be entangled even though space is no longer connected.
 
 \setcounter{equation}{0}
\section{Further Applications}
In this section we discuss applications of tachyon condensation to black $p$-branes, asymptotically AdS solutions, and extremal black holes.
In some of these cases, the result of the transition was anticipated in the literature, but the stringy mechanism behind it was not understood. 

\subsection{Black $p$-branes}

So far we have focused on black strings, but similar arguments apply to higher dimensional branes\footnote{To obtain a solution to ten dimensional string theory, one must take a product of the $D$ dimensional black strings discussed in the previous section and a $10-D$ dimensional solution. So they can be viewed as smeared black branes. We now consider the more fundamental (nonsmeared) branes.}.  Consider a black p-brane with RR charge
 \be\label{p-brane}
  ds^2 = H_p^{-1/2}(r) [-f(r) dt^2 + dx_i dx^i] + H_p^{1/2} (r)[ f^{-1}(r) dr^2 + r^2 d\Omega_{n+1}] 
  \ee
  where $n=7-p$, $i=1,\cdots, p$,  and
  \be 
 H_p(r) = 1+ {r_0^{n}\sinh^2\al\over r^{n}}, \qquad f(r) = 1-{r_0^n\over r^n}
 \ee
 The RR field and dilaton are
 \be
 C_{01\cdots p} =  {r_0^n\sinh2\al \over 2(r^n+r_0^n\sinh^2 \al)}, \qquad e^{\Phi} = H_p^{(3-p)/4}(r)
 \ee
If we wrap this brane around a circle by periodically identifying one of the $x^i$ with period  $L$, then this circle again shrinks to zero size at the singularity. If there are antiperiodic fermions around this circle, there will again be tachyon condensation when its size reaches the string scale.  For $\cosh^{1/2}\al = L/l_s$, this will occur at the horizon, and the result is a  KK bubble with  RR charge.  Static bubbles of this type can be obtained by a double analytic continuation $t\ra i\chi,\ x_1 \ra it$ from (\ref{p-brane}).
However, just like the single charged black string in section 2, most of these static bubbles cannot be  the end state  of tachyon condensation. 
The periodicity of the $\chi$ circle is again given by (\ref{period}). Setting this equal to $l_s\cosh^{1/2}\al$ yields $r_0 \cosh^{1/2}\al \approx l_s$ which implies that the curvature at the horizon is of order the string scale. This is the correspondence point
\cite{Horowitz:1996nw} when the description of the black brane should change to open  strings on D-branes. However, it is easy to arrange for the black brane to undergo tachyon condensation when the curvature at the horizon is much smaller than the string scale. (In particular, this is always true if $r_0 > l_s$.) The resulting bubble must expand out.

The one exception is the three brane. This is analogous to the two-charged black string since one can allow the parameters of the bubble, $\tilde r_0$ and $\tilde \al$ to differ from the black brane.  The condition for the size of the horizon to equal the size of the bubble, and the charge to be conserved are both satisfied (for large $\al,\tilde\al$) if $r_0^2 e^\al = \tilde r_0^2 e^{\tilde\al}$. The constraint on  forming  static bubbles now depends on the charge and $L$. Setting $n=4$ in (\ref{period}) and
$Q= r_0^4\sinh 2\al$ one finds 
\be
{Q\over L^4} \propto {\sinh\al\over \cosh^3\al}
\ee
which is bounded from above (and, surprisingly, the same as the constraint arising from the two charged black string (\ref{twocharged})).
 
\subsection{Asymptotically AdS solutions}

 As is well known, the near horizon limit of the extremal three-brane is $AdS_5\times S^5$. If we periodically identify one direction in Poincare coordinates,  this spacetime has an orbifold-like singularity at the horizon. It was shown in \cite{Horowitz:1998ha} that with antiperiodic boundary conditions for the fermions around this circle, there is a lower energy static solution, called the AdS soliton:
\be
ds^2 = {r^2\over \ell^2}\[ -dt^2 + f(r) d\chi^2 +dx^2 + dy^2\] + f^{-1}(r) {\ell^2\over r^2} dr^2
\ee
 This solution is just the near horizon limit of the static $p=3$ bubble mentioned above. It was shown that this solution is  perturbatively stable, and its  energy   agrees with the negative Casimir energy of the dual CFT compactified on a circle with SUSY broken by the same boundary conditions\footnote{More precisely, the Casimir energy can only be computed at weak coupling and agrees with the supergravity energy up to a factor of $3/4$, the same factor which arises when comparing the near extremal entropy.}. Based on this, the AdS soliton was conjectured to be the lowest energy state on the gravity side with these boundary conditions, and there is growing evidence that this is the case \cite{Galloway:2002ai,Anderson:2004yi}.  It was reasonable to expect that some transition would cause periodically identified $AdS_5$ to decay to this lower energy ground state, but no mechanism was known. We now see that tachyon condensation will do precisely this. 
 
 It is interesting to compare this example with the case of $AdS_3\times S^3$ above. Even though the geometry of the static bubble (AdS soliton) is now different from global $AdS_5$, one still finds a stable ground state. It appears that asymptotically AdS boundary conditions  stabilize bubble type geometries, and avoid the catastrophic consequences that arise in the asymptotically flat context. However, in both of these examples, the asymptotic circle was one of the directions in Poincare coordinates. There are other possibilities where stability is less clear.

 For example, $AdS_5$ can be written
\be\label{5dbtz}
ds^2 = \(1+{r^2\over \ell^2}\) dx^2 + \(1+{r^2\over \ell^2}\)^{-1} dr^2 + r^2[-dt^2 + \cosh^2 t d\Omega_2]
\ee
If $x$ is periodically identified, this spacetime is reminiscent of a black hole with horizon at $r=0$.  It is a higher dimensional analog of the BTZ black hole \cite{Banados:1997df}, and the boundary at infinity is $S^1\times dS_3$. The $x$ circle reaches a minimum size at $r=0$. If this size is the string scale (and with antiperiodic fermions), tachyon condensation will cause it to pinch off. The result is not global $AdS_5$, even though the resulting space now has the same global topology as $AdS_5$. The reason is that  (\ref{5dbtz}) has negative energy  \cite{Balasubramanian:2002am,Cai:2002mr}. Tachyon condensation cannot increase the mass.  Instead, it will produce a  bubble solution. One example of a bubble with the right asymptotic geometry and lower mass is the double analytic continuation of the five dimensional Schwarzschild AdS geometry \cite{Balasubramanian:2002am,Birmingham:2002st}. In fact, 
 in analogy with the original Kaluza-Klein vacuum instability,  it was argued in \cite{Balasubramanian:2005bg}   that the space (\ref{5dbtz}) could decay via a gravitational instanton to this bubble of nothing. We now see that for string scale circles, there is a much more rapid  mechanism for this decay.

\subsection{Extremal black holes}

Our final application involves an old argument due to Hawking for why black holes have an intrinsic entropy while traditional solitons do not \cite{Hawking}. If one analytically continues a static soliton and periodically identifies euclidean time with period $\beta$, the action is simply $I = \beta H$ where $H$ is the Hamiltonian. The usual thermodynamic formula for the entropy is
\be
S =\(1- \beta {\partial\over \partial \beta} \) \ln Z
\ee
where the partition function $Z$ is given by the integral of $e^{-I}$ over all configurations which are periodic in imaginary time with period $\beta$. If we approximate the partition function by the classical saddle point, $Z=e^{-I}$, we clearly get $S=0$. If one applies this to a black hole, one gets an extra term in the action coming from the fact that the euclidean solution is not a product of space and (euclidean) time. The constant $t$ surfaces all meet at the horizon. One can write the action as a sum of a contribution from a small neighborhood of the horizon and everything outside.  One finds $I_{bh} = \beta H - A_{bh}/4$, which yields the familiar result $S_{bh} = A_{bh}/4$.  However, an extremal black hole is different.  
 The horizon is infinitely far away, and the euclidean black hole is topologically a product. This led to the suggestion that extremal black holes might have zero entropy \cite{Hawking:1994ii}. Shortly after this prediction was made, Strominger and Vafa \cite{Strominger:1996sh} calculated the entropy of an extremal black hole in string theory and found the familiar result $A_{bh}/4$.  It was suggested in \cite{Horowitz:1996qd} that one way to reconcile the euclidean quantum gravity argument with the string theory calculation is if string theory  modifies the euclidean geometry for an extremal black hole. In particular, the euclidean time circles shrink to zero size as one approaches the horizon and finite temperature boundary conditions require antiperiodic fermions, so the winding modes become tachyonic.  The outcome of the tachyon condensation was not known at the time, but it was suggested that IF the tachyon leads to a modified solution with the same topology as the nonextremal euclidean black hole, then the euclidean arguments would  reproduce the usual answer: $S_{bh}=A_{bh}/4$.  We now see that this is exactly what happens.

\setcounter{equation}{0}
\section{Discussion}
We have seen that in the course of Hawking evaporation, closed string tachyons can cause a black string or black brane to turn into a Kaluza-Klein bubble of nothing. Depending on the charge and the size of the wrapped circle, the resulting bubble will be static or expanding. With asymptotically Kaluza-Klein boundary conditions, the nonstatic bubbles rapidly expand out to infinity cutting off all space. This is certainly a dramatic endpoint of Hawking evaporation. However the antiperiodic boundary conditions
required for the fermions produce a Casimir energy which destabilizes the asymptotic circle. More physical applications require that the circle be stabilized as in \cite{Saltman:2004jh} (or perhaps in cosmology) and the bubble evolution in these contexts remains to be investigated.

We have focused on  charged black strings, but it is clear that  similar results hold in the extremal limit. If one wraps a sufficient number of branes around a circle (with the right spin structure) the $S^1$ will become string scale when the curvature is still small. For the strings in section 2, this will be true if the winding number, $N$, and string coupling, $g$, satisfy $g^2 N > (L/l_s)^2$. The result is  again  a bubble. One can also take a neutral black hole and throw in a large number of  wrapped strings (or D-branes) and induce the transition.

We have discussed black strings with two charges, but  what about the famous three charged strings which reduce to extremal five dimensional black holes? 
It is straightforward to add momentum to the above black strings by applying a boost: $t\ra t\cosh\beta + x\sinh\beta$, $x\ra x\cosh\beta +t\sinh\beta$.  Since this results in 
\be
g_{xx} = {r^2+r_0^2\sinh^2\beta\over r^2 + r_0^2 \sinh^2\al}
\ee
the circle now has a minimum radius. However if the winding charge is much greater than the momentum, $\beta \ll \al$, the circle can still reach the string scale at the horizon. Tachyon condensation will again produce a  bubble. This is puzzling since a bubble cannot carry momentum. The identification which produces a smooth geometry must be done in the rest frame of the bubble. This follows from the fact that the identification is always along a Killing field which vanishes at the bubble, which is the original (unboosted) $\partial/\partial \chi$.  The resolution of the puzzle seems to be that the momentum of the black string must be carried by the radiation produced when the bubble is formed. This can also be seen in the near horizon limit. A (near extremal) three charge black string has a near horizon limit which is a rotating BTZ black hole. If the size of the horizon is of order the string scale, there will be a tachyon  transition to global $AdS_3$ which has zero angular momentum. The angular momentum remains in the radiation produced.

One can also add rotation to the charged black strings and black branes without affecting much of the results.  Although we have not studied this in detail, it seems clear that tachyon condensation will still produce  bubbles of nothing. However, the analog of the static bubble will now have  ``twisted circle" boundary conditions at infinity where space is identified  under a translation plus rotation. These circles open up at infinity, so supersymmetry is locally preserved. Neutral bubbles in such spacetimes have been studied in \cite{Dowker:1995gb}.

As mentioned in the introduction, understanding tachyon condensation near the horizon of a black string is only the first step toward a complete description of tachyon condensation in the evaporation process. Significant progress has recently been made in understanding tachyon condensation inside the horizon \cite{MS}, and the remaining task  of matching the interior description to the exterior description 
 is currently under investigation \cite{HMS}. 
\vskip 1cm
\centerline{\bf Acknowledgments}
\vskip .5cm
It is a pleasure to thank H. Elvang and S. Minwalla for discussions, and  especially J. McGreevy and E. Silverstein for very helpful  correspondence and collaboration at an early stage of this project. This 
work was supported in part by
NSF grant PHY-0244764.

\end{document}